\documentclass[reprint,amsmath,amssymb,aps,prl,groupedaddress,nofootinbib,superscriptaddress]{revtex4-1}
\usepackage{graphicx}
\usepackage{amsthm,amssymb,amsmath,braket,mathdots}
\usepackage{bm}
\usepackage[hidelinks,pagebackref=false,pdfnewwindow=true]{hyperref} 
\usepackage{epstopdf,psfrag}
\usepackage{relsize,amsbsy}
\usepackage[export]{adjustbox}

\usepackage{graphicx,xcolor,tikz}

\newcommand{\be}{\begin{equation}}
\newcommand{\ee}{\end{equation}}

\newcommand{\mss}{Majorana spinons }

\newcommand{\bit}{\begin{enumerate}}
	\newcommand{\eit}{\end{enumerate}}

\definecolor{bananayellow}{rgb}{1.0, 0.88, 0.21}
\definecolor{straw}{rgb}{0.32, 0.28, 0.1}

\definecolor{palatinatepurple}{rgb}{0.49, 0.24, 0.46}

\definecolor{darkgreen}{rgb}{0.0, 0.5, 0.0}

\def\eg{\emph{e.g.}\ }

\newcommand{\pa}{\partial}

\graphicspath{{Figures/}}

\begin{document}

	\title{Majorana Landau Level Raman Spectroscopy}
		\author{Brent Perreault}
		\affiliation{\small School of Physics and Astronomy, University of Minnesota, Minneapolis, Minnesota 55455, USA}
		\author{Stephan Rachel}
		\affiliation{\small Institut f\"ur Theoretische Physik, Technische Universit\"at Dresden, 01062 Dresden, Germany}
		\affiliation{\small Department of Physics, Princeton University, Princeton, New Jersey 08544, USA}
		\author{F. J. Burnell}
		\affiliation{\small School of Physics and Astronomy, University of Minnesota, Minneapolis, Minnesota 55455, USA}
		\author{Johannes Knolle}
		\affiliation{\small Department of Physics, Cavendish Laboratory, JJ Thomson Avenue, Cambridge CB3 0HE, U.K.}


\begin{abstract}
The unambiguous experimental detection of quantum spin liquids and, in particular, of the long-sought Kitaev quantum spin liquid (KQSL) with its Majorana fermion excitations remains an outstanding challenge.
One of the major obstacles is the absence of signatures that definitively characterize this phase. 
Here we propose the Landau levels known to form in the Majorana excitation spectrum of the KQSL when certain strain fields are applied as a direct signature of Majorana fermions with Dirac-like dispersion.  In particular, we
show that the Majorana Landau level quantization of strained films of the KQSL can be directly probed by Raman spectroscopy.  Such experiments are feasible in thin films 
of $\alpha$-RuCl$_3$, which are a promising place to search for the KQSL.
\end{abstract}
		
		
		\maketitle
		

	
	
\paragraph{Introduction.}
Two of the most prominent -- but normally distinct -- routes for the experimental realization of topologically ordered (TO) phases of matter are 2D electron gases in strong magnetic fields and frustrated magnets. In the former, the orbital magnetic field leads to Landau quantization, introducing an extensive number of degenerate single-particle states.  In the presence of interactions, this leads to TO fractional quantum Hall (FQH) phases. In the latter, geometrically frustrated interactions between the effective magnetic moments lead to a large classical degeneracy, which can result in so-called quantum spin liquid (QSL) phases at low temperature.  
Strain engineering of certain thin QSL films unifies these distinct lines of research by inducing pseudo-magnetic fields~\cite{Guinea10} for the fractionalized, chargeless quasiparticles of QSL phases~\cite{Rachel15}.  
In this work, we show that for at least one type of spin liquid the resulting emergent Landau quantization can be detected experimentally with inelastic light scattering.  
 This allows both identification of the spin liquid phase and characterization of the band structure of its fractionalized fermionic excitations, which have thus far proven experimentally elusive.

\begin{figure}[t!]
	\centering
	\includegraphics[width=0.9\linewidth]{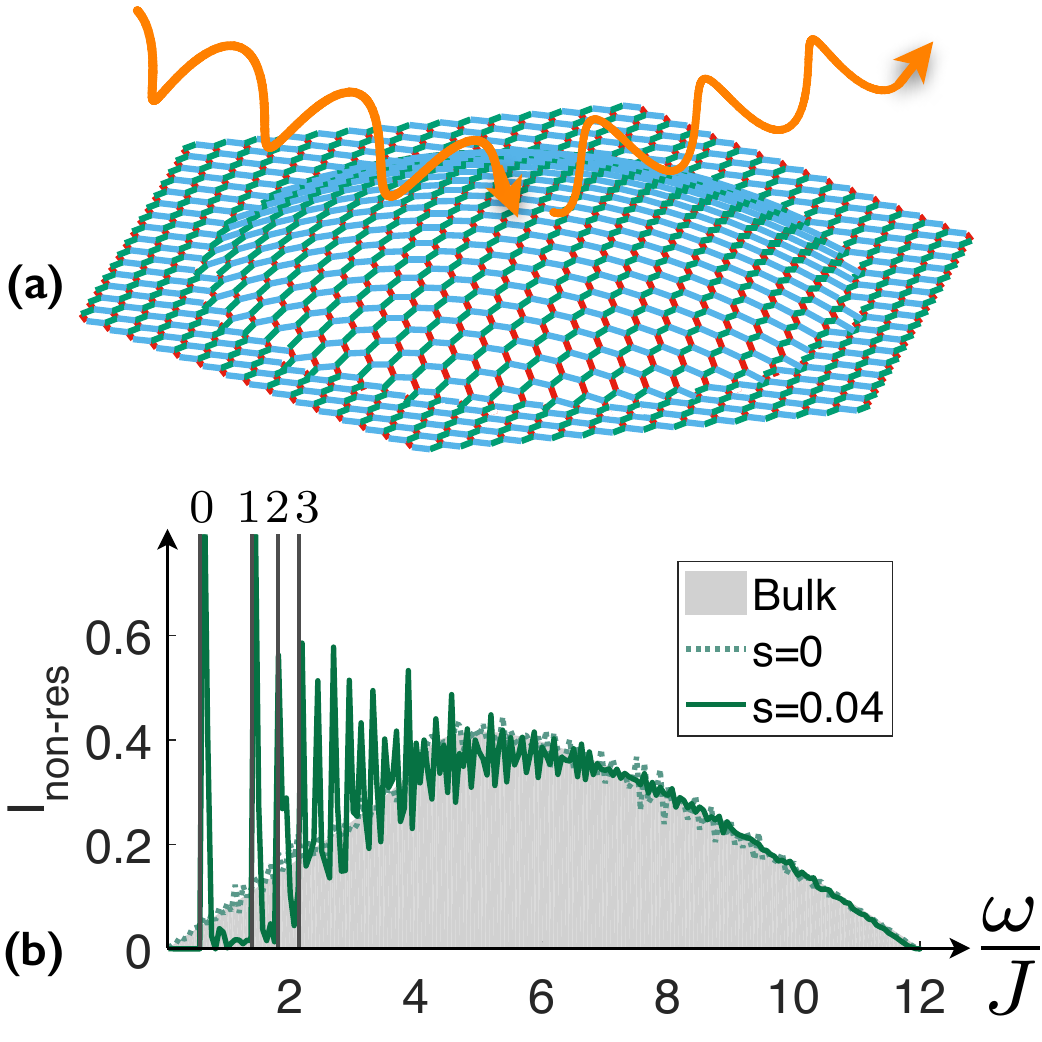}
		\caption{Schematic plot of a nano-bubble of a $\alpha$-RuCl$_3$ thin film is shown in panel (a). The resulting lattice distortion induces a strain pattern that acts as a pseudo-orbital magnetic field for the Majorana fermions which emerge from spin fractionalization in the QSL of the Kitaev model. The resulting LL quantization can be directly probed by Raman scattering. The evolution of the response for finite $(s=0.04)$ and zero strain $(s=0)$ is shown in panel (b).}
	\label{Fig1}
\end{figure}
	
Our study focuses on the so called Kitaev QSL (KQSL), which is the ground state of an exactly solvable spin-$\frac{1}{2}$ model with frustrated bond-dependent Ising  interactions on the honeycomb lattice~\cite{Kitaev06}. There spins fractionalize into emergent static Z$_2$ fluxes and dispersive Majorana fermions which at low temperatures (corresponding to few excited fluxes) display a linear Dirac spectrum similar to graphene. The exact solubility makes this model particularly amenable to theoretical study; indeed both static properties (\eg ground state degeneracy\,\cite{Mandal2012}, entanglement entropy\,\cite{Yao2010}, and disorder\,\cite{Willans2010,Lahtinen2014}) and dynamical correlations (e.g.\ dynamical structure factor\,\cite{Knolle2014,Knolle2015}, Raman scattering\,\cite{Knolle2014b,Perreault2015}, and global quenches\,\cite{Sengupta2008}) of this model can be computed. 

Exact solubility also allows the effect of strain on the KQSL to be characterized analytically.  It is well known that lattice strain couples to electronic degrees of freedom as an effective gauge field\,\cite{suzuura-02prb235412,Manes2007} which has led to the prediction\,\cite{Guinea10} that suitably strained graphene would display the characteristic energy scaling $E_n \propto \text{sgn}(n) \sqrt{|n|}$ of Dirac electrons in an orbital magnetic field. This was successfully demonstrated in subsequent experiments\,\cite{Levy10} where the strain pattern was 
generated by nano-bubbles which can form when graphene is grown on a suitable substrate.  	
Recently, it was shown theoretically that the Majorana fermions in a strained Kitaev model also experience pseudo-magnetic fields\,\cite{Rachel15}, leading to Landau level (LL) quantization with the same energy scaling as for graphene.   

The prediction that certain transition metal oxides with strong spin orbit coupling and orbital degeneracies could be dominated by Kitaev spin-exchange interactions \cite{Jackeli09} has recently inspired a search for candidate KQSL materials.  Initially, interest was focused on A$_2$IrO$_3$ (A$=$Na, Li)\,\cite{Singh2012,Comin2012,Biffin2014,Takayama2015} but currently $\alpha$-RuCl$_3$\,\cite{Plumb2014,Kubota2015} appears to be the best KQSL candidate. Though all of these compounds show magnetic order at low temperature\,\cite{Schaffer2016}, it has been argued that dynamical scattering experiments --  inelastic neutron scattering (INS)\,\cite{Banerjee2016,Banerjee2016c} and Raman scattering\,\cite{Sandilands2015,Nasu16}  -- can, at higher frequencies, naturally be interpreted in terms of the fractionalized Majorana fermions\,\cite{lowfreq}. This suggests that $\alpha$-RuCl$_3$ may be proximate to the KQSL, such that tuning material parameters could potentially drive the system into a QSL.   Further, strain engineering in these systems is conceivable in the near future: As the material is highly two dimensional, thin films can be generated by simple exfoliation techniques\,\cite{ziatdinov-17}. 
Strain could therefore be generated by  placing these on different substrates or shapes. 

This raises the exciting possibility of detecting KQSL physics by probing the LL quantization of the deconfined Majorana fermions.  This would give a ``smoking gun" signature of the existence of Majorana fermions {\it with a linear Dirac spectrum} characteristic of the KQSL.
However, it is not obvious how to detect the Majorana LLs.  Since these fermions are charge neutral, the LLs cannot be detected simply by means of scanning tunneling microscopy as for graphene~\cite{Levy10}. Further, probes such as INS or RIXS cannot produce adequate signal on very thin films and do not have the necessary spatial resolution. Here, we show that inelastic light scattering of strained Kitaev films would probe the Majorana LL quantization and the characteristic scaling $E_n \propto +\sqrt{|n|}$ related to the Dirac dispersion. 


\paragraph{Low-energy description and scaling.}
The Kitaev honeycomb spin model is given by~\cite{Kitaev06}
\begin{align}\label{H}
{H}_K &= \sum_{\left<ij\right>^\alpha} J^\alpha \sigma^\alpha_i \sigma^\alpha_j ,
\end{align}
with only one spin component $\alpha$ interacting on each of the three distinct bonds of the honeycomb lattice.  Kitaev's solution decomposes spins into products of Majorana fermions via $\sigma^\alpha_j = i b^\alpha_j c_j$, such that
${H}_K = \sum_{\left<ij\right>^\alpha} J^\alpha u_{\left<ij\right>^\alpha} i c_i c_j$
where $u_{\left<ij\right>^\alpha} = i b_i^\alpha b_j^\alpha$ are conserved $\mathbb{Z}_2$ gauge variables. Since the gauge fluxes are gapped and static we can first focus on the zero-flux ground-state configuration, \eg with $u_{\left<ij\right>^\alpha} = 1$. This reduces the problem to that of a quadratic Hamiltonian of the dispersing $c$ Majorana fermions. For isotropic exchange couplings and no strain it gives a linear spectrum $\epsilon_k = 2|\Gamma_k|$, with $\Gamma_{\mathbf k} = J^z + J^x e^{ik_1} + J^y e^{ik_2} $ and $k_i = {\mathbf k}\cdot{\mathbf a}_i = (\pm \sqrt{3}k_x + 3k_y)/2$,  about the two Dirac points $k_1 = -k_2 = \pm 2\pi/3$ similar to graphene.

A strain field modifies the parameters $J^\alpha$ in Eq.\,(\ref{H}). At lowest order 
the correction to the unstrained exchange couplings
 is given by\,\cite{Rachel15} 
\begin{align}
\delta J_{ij}^\alpha / J^\alpha &= - \beta \left( |\vec{\delta}_{ij}| - 1 \right) \approx -\beta (\vec{\delta}_0 \cdot \vec{\nabla})(\vec{U}\cdot \vec{\delta}_0) , 
\end{align}
where $\delta_0$ is the lattice vector in the absence of strain,  $\vec{U}_j = \vec{R}_j' - \vec{R}_j$ is the displacement field, and $\vec{\delta}_{ij} = \vec{R}_i' - \vec{R}_j'$ is the strained lattice vector. Expanding the Hamiltonian around the Dirac points to first order in strain and wave vector it is easy to show that it couples just like a vector potential $\mathbf{\Pi}= \mathbf{p}-\frac{e}{c} \mathbf{A}$ to the canonical momentum. An out of plane  magnetic field (along the $z$-axis) is induced by strain fields of the form $B=-\beta \left[\partial_x u_{xy} + \frac{1}{2} \partial_y (u_{xx}-u_{yy}) \right]$~\cite{Guinea10} and $\beta \!\equiv\! -\frac{\partial \ln J}{\partial \ln \delta}$ is the magnetic Gr\"uneisen parameter \cite{White65}.

\begin{figure*}
	\centering
	\includegraphics[width=\linewidth]{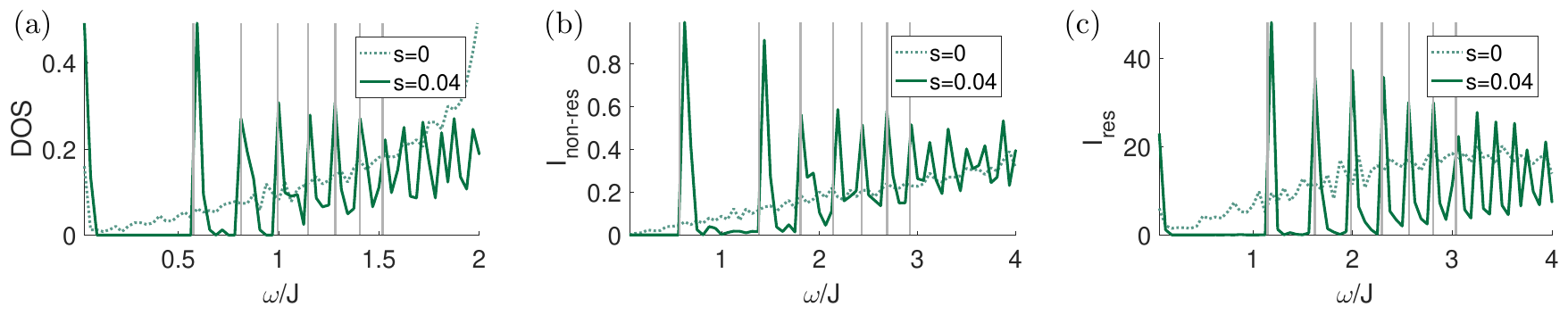}
	\caption{The DOS and all possible Raman correlation functions for finite strain ($s=0.04$) [solid lines] and zero strain ($s=0$) [dotted lines] with $r=50$ corresponding to $6*(50+1)^2 = 15606$ sites. DOS (a) shows peaks at $\sqrt{n}\omega_c$ (gray vertical lines) while $I_{\text{non-res}}$ (b) features them at $(\sqrt{n} + \sqrt{n+1})\omega_c$ and $I_{\text{res}}$ (c) at $2 \sqrt{n} \omega_c$. 
		$I_{\text{res}}$ is the resonant channel, which is anti-symmetric in polarization, as discussed in Ref.\,\cite{Perreault16-1}.
%
}
	\label{basicPlots}
\end{figure*}

Before presenting numerical results for the lattice model, we give an analytical treatment of the low-energy physics.  We work in the Landau gauge $\mathbf{A}=B(0,x)$ and introduce the low-energy Majorana field operators  $\hat \Psi_{\nu} = \left( \Psi_{a}(\mathbf{r}), \Psi_{b}(\mathbf{r}) \right)^T$. The index $\nu=\pm 1$ labels the two valleys while $a$ and $b$ refer to the two sublattices.
Now we can write 
 the Hamiltonian as $H = \sum_{\nu} \int \text{d}^2 \mathbf{r} 
\hat \Psi^{\dagger}_{\nu}
\hat H_{\nu}
\hat \Psi_{\nu} $ with
\begin{eqnarray}
\hat H_{\nu} = i \frac{3}{2} J 
\begin{pmatrix}
0 & \nu p_x- i(p_y- \nu B x) \\
- \nu p_x- i(p_y-\nu B x) & 0 
\end{pmatrix} .\nonumber
\end{eqnarray}
Note that $B$ has opposite sign at the two Dirac points leaving time reversal symmetry (TRS) unbroken, which prevents the usual trick of combining the two Majorana cones into a single cone of complex fermions~\cite{Balents2016}. As long as there is no coupling between the two cones we can concentrate on only one of them.
We introduce the ladder operator $a=\frac{l_B}{\sqrt{2}\hbar} \left(\Pi_x-i \Pi_y \right)$ with $l_B^2=\frac{c\hbar}{e|B|}$ (in the following we absorb $e,c$ into the definition of $B$ and set $\hbar=1$) and expand the field operators in terms of the standard Landau Level wave functions $\Psi_{a} (\mathbf{r}) = \frac{1}{\sqrt{2}} \sum_{n,p} \Phi_{n-1,p} (\mathbf{r}) c_{a,n,p}$ and $\Psi_{b} (\mathbf{r}) = \frac{1}{\sqrt{2}} \sum_{n,p} \Phi_{n,p} (\mathbf{r}) c_{b,n,p}$, where
the Majorana operators $c_{X,n,p}$ ($X=a,b$) anticommute, with $c^2 = 1$ [note the sign change of momenta from conjugation $\Psi_{b}^{\dagger} (\mathbf{r}) = \sum_{n,p} \Phi^*_{n,p} (\mathbf{r}) c_{b,n,-p}$]. 
Using the standard properties of ladder operators, $a \Phi_n=\sqrt{n} \Phi_{n-1}$ and $a^{\dagger} \Phi_n=\sqrt{n+1} \Phi_{n+1}$, we obtain the Hamiltonian 
\begin{eqnarray}
H_{+}  \!& = & \!\frac{3J}{2} \sum_{n,p}
\mathbf{c}^T_{n,-p}
\begin{pmatrix}
0 & i \frac{\sqrt{2}}{ l_B} \sqrt{n} \\[5pt]
- i \frac{\sqrt{2} }{ l_B} \sqrt{n} & 0 
\end{pmatrix}\!
\mathbf{c}_{n,p},
\end{eqnarray}
where $\mathbf{c}^T_{n,p} = (c_{a, n, p}, \ c_{b, n, p})$.
We diagonalize $H_+$ by the complex fermions $f_{n,p}$ with $c_{b,n,p}=f_{n,p}+f^{\dagger}_{n,-p}$ and $c_{a,n,p}=i \left(f_{n,p}  -f_{n,-p}^{\dagger}\right)$ such that
$H_{+}  =  \sum_{n,p} E(n) \left[ f_{n,p}^{\dagger} f_{n,p} -\frac{1}{2}\right] $. The energies  \cite{Rammal85}
\begin{eqnarray}
E(n)= \omega_c \sqrt{n}
\end{eqnarray}
obey the well known $\sqrt{n}$ scaling of Dirac fermions with $\omega_c= \frac{3  J\sqrt{2}}{ l_B} = 3\sqrt{2} J \sqrt{B}$ and $n \in \mathbb{N}_{\geq 0}$.

We are interested in the low-energy behavior of the Raman response which has been discussed for unstrained Kitaev models by some of us in the past~\cite{Knolle2014b,Perreault2015,Perreault16-1,Perreault16-2}. 
The Raman intensity is given by the correlation function
 $I(\omega)=\int_{-\infty}^{\infty} \text{d}t e^{i\omega t} \langle R(t) R(0)\rangle$, where the effective Raman vertices $R(t)$  depend on the in- and out-going polarization of the scattered photons. We concentrate on strain patterns that do not alter the symmetries of the Hamiltonian such that there are only two independent Raman intensities, the $A_{1g}=E_{g}$ channel and the $A_{2g}$ channel which only couples to incident photons in resonance with the minimal Mott gap.
 The main difference between the non-resonant and resonant Raman vertices is that the latter can couple sites on the same  sublattice, whereas the former cannot~\cite{Perreault16-1}.  As we are only interested in the scaling form of the Raman response we omit the polarization dependent prefactors to obtain
\begin{align*}
R_{\text{non-res}}  &\propto  i \int \text{d}^2 \mathbf{r}  \Psi^{\dagger}_a(\mathbf{r})(t) \Psi_b(\mathbf{r}) \\
&\propto  \sum_{n,p} \left[ f_{n,-p}(t) -f^{\dagger}_{n,p} (t) \right] \left[ f_{n-1,p} + f^{\dagger}_{n-1,-p} \right]
\end{align*} 
for the non-resonant processes, and
\begin{align*}
R_{\text{res}} &\propto i \int \text{d}^2 \mathbf{r} \left[ \Psi^{\dagger}_a(\mathbf{r})(t) \Psi_a(\mathbf{r+\delta}) - \Psi^{\dagger}_a(\mathbf{r})(t) \Psi_a(\mathbf{r-\delta}) \right] \\ 
 &\propto  \sum_{n,p} \sin \left(p \delta \right) \left[ f_{n,-p}(t) -f^{\dagger}_{n,p} (t) \right] \left[ f_{n,p} + f^{\dagger}_{n,-p} \right] 
 \end{align*} 
for the antisymmetric combination of polarizations in the resonant processes.
Only the non-resonant combination mixes states which differ by one LL index. From the time dependence $f_{n,p}(t)=f_{n,p} e^{-i t E(n)}$ and $f_{n,p} |0\rangle=0$ we can directly calculate the low-energy Raman responses,
\begin{equation}\begin{split}
\label{Scaling}
I_{\text{non-res}} & \propto   \sum_{n} \delta\left[\omega -  \omega_c \sqrt{n}- \omega_c \sqrt{n+1} \right]\ , \\ 
I_{\text{res}} & \propto   \sum_{n} \delta\left[\omega - 2 \omega_c \sqrt{n}  \right] .
\end{split}\end{equation}
This is the central result of the paper: 
the Raman response \eqref{Scaling} is a direct probe of the LL quantization. 
The two different scalings of the resonant and non-resonant intensities in Eq.\,\eqref{Scaling} originate from the sub-lattice selectivity of the two vertices. 

\begin{figure*}
	\centering
	\includegraphics[width=\linewidth]{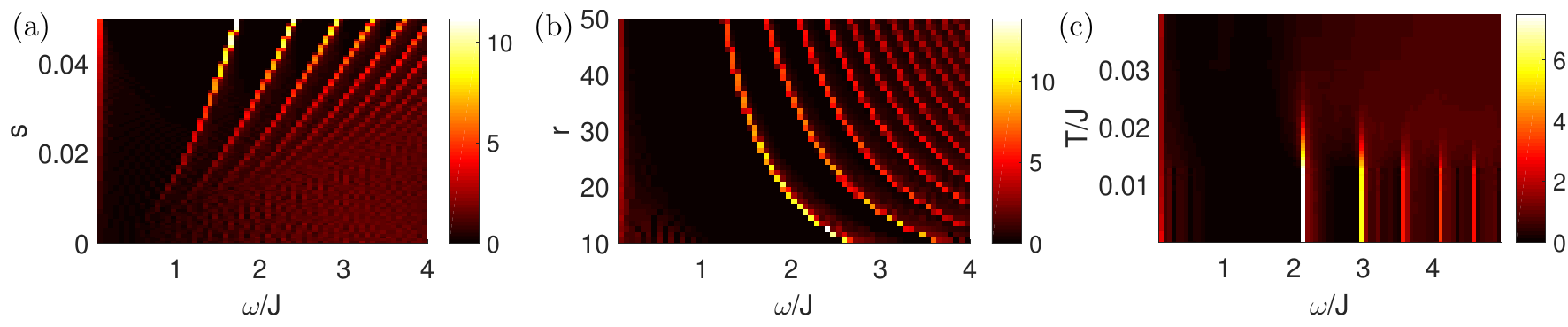}
	\caption{A set of density plots showing the progression of Landau level peaks as a function of (a) strain $s$, (b) system size $r$, and (c) temperature $T$. Where not specified we used parameters $s=0.04$ and $T=0$ with the system sizes (a) $r = 30$ and (c) $r = 15$. The temperature of the crossover in (c) agrees well with the expected spinon-confinement transition for non-interacting fluxes at $T^* \approx 0.02J$ ($s=0.04$, $r=15$).} 
	\label{colorPlots}
\end{figure*}

\paragraph{Lattice calculation.} 
We follow Refs.~\cite{Guinea10,Amal13,Rachel15} to study a honeycomb flake subject to triaxial strain preserving the $C_{3v}$ symmetry. 
The strain pattern defined by the displacement field 
\begin{align}
\vec{U}(x,y) = \bar{C}\big(2xy,x^2-y^2\big)
\end{align}
leads via $u_{ij} = (\pa_i U_j + \pa_j U_i)/2$
 to a uniform  pseudo-magnetic field with magnitude $B = -4 \beta \bar{C}$. 
We numerically construct the honeycomb flake as $r$ rings of honeycombs placed around an initial single one\,\cite{Rachel15,conventions}. We let $\bar{C} = \bar{C}(r)$ depend on system size so that the fractional stretch in the maximal direction 
$s = \frac{\delta L}{L} = \sqrt{3}(r+\frac{1}{2})\bar{C}$ is fixed. Then $B \approx -\frac{4\beta}{\sqrt{3}r}s$ decays with system size. This also fixes the maximum magnetic response: For linear elasticity to hold $\delta J / J \ll 1$, or $s \ll \frac{\sqrt{3}}{2\beta}$ independent of system size. 

For concreteness we use $\beta = 10$ which gives for optimal parameters strain of $s=0.01$--$0.04$ in systems  of flake size $r = 10$--$50$ unit cells.  
The pseudo-magnetic field needs to be sufficiently strong that the Landau levels are easily distinguished from finite-size effects in the Raman response; for these system sizes this requires at least $B\approx 0.03$ or strains of $s=0.02$ if $\beta=10$. 

In Fig.\,\ref{Fig1}\,(b) we show how the characteristic Raman response for an unstrained thermodynamically large system transforms under strain. Without strain the linear slope at low energies from the Dirac density of states (DOS)~\cite{Knolle2014b} is apparent. 
The Figure also shows that the non-resonant response of a strained flake obeys the characteristic discrete scaling predicted by our low-energy theory [Eq.\eqref{Scaling}], which is a direct consequence of the emergence of Landau quantization. 
In Fig.\,\ref{basicPlots} we concentrate on the low-frequency window and compare the evolution of the Majorana fermion DOS [Fig.\,\ref{basicPlots}\,(a)], non-resonant [Fig.\,\ref{basicPlots}\,(b)] and resonant Raman response [Fig.\,\ref{basicPlots}\,(c)]. We recover the distinct scaling of the resonant and non-resonant processes \eqref{Scaling}. 

In order to make contact with possible future experiments, Fig.\,\ref{colorPlots} traces the evolution of the LL peaks as a function of strain and flake size in panels (a) and (b). 
They follow the expected behaviour of the effective magnetic field $|B| \approx \frac{4\beta}{\sqrt{3}r}s$. Panel (c) shows the evolution with temperature. 
For a fixed finite system size, there is a cross-over temperature $T^*>0$ above which the fluxes destroy the low-energy Dirac spectrum, washing out the sharp LL quantization.   A uniform thermal density $\rho$ of fluxes effectively confines the \mss on a length scale $\ell_C \sim \rho^{-1/2}$; it is therefore natural to expect that when $\ell_C$ approaches the system's effective magnetic length the Landau levels dissappear.  However, as the strain grows with the radial coordinate, there turns out to be significant spatial anisotropy in the flux gap \cite{Rachel15} causing the fluxes to first appear near the edge of the system (see the supplemental material\,\cite{supp}). This allows the Landau levels to persist to an appreciable flux density of around $\rho \approx 0.1$ at $s = 0.04$. 
For a fixed value of 
strain $s$ and the magnetic response $\beta$ the spatial flux anisotropy is expected to be independent of system size, following the spatial strain anisotropy. We can therefore predict a crossover temperature of around $T^* \approx 0.02J$ for any system size with this strain pattern at $s=0.04$ and $\beta = 10$ based on our numerical results. 
Details of the finite temperature calculation, including comparison to known results\,\cite{Nasu15,Nasu16}, are relegated to the supplementary material\,\cite{supp}.


{\it Discussion.}
Our calculation of the Raman response in a strained KQSL shows that it can be used to directly observe the LL quantization and the corresponding low-energy scaling.  This gives direct evidence for the Dirac dispersion of the Majorana fermions in the KQSL.   This distinctive signature is remarkable because in general it is notoriously difficult to measure asymptotic low temperature (or low frequency) properties that can definitively identify the QSL type in candidate materials. 

We emphasize that this LL quantization -- which produces a distinctive signature in the Raman response up to energy scales that are an appreciable fraction of the bandwidth (see Fig. \ref{Fig1}) --  is a feature of the low-temperature crossover region $T<T^*$ proximate to the zero temperature quantum spin liquid phase. This is in contrast to the broad continuum 
observed in unstrained $\alpha$-RuCl$_3$~\cite{Sandilands2015}, which compares favorably with the predictions for the Kitaev model even in samples with low-temperature magnetic order.  For pure Kitaev interactions such a continuum is present even at relatively high temperatures $T \gg T^*$ throughout the correlated paramagnetic regime up to a temperature set by the value of the Kitaev exchange~\cite{Nasu16}.  
In contrast, the sensitivity of the LL degeneracy to flux disorder means that it disappears at temperatures where fluxes proliferate in the bulk and destroy the low-energy Dirac dispersion of the spin liquid.
 
In order for these Landau levels to be observable, it is crucial that the LL quantization is a feature of the KQSL phase, rather than of the finely-tuned Kitaev Hamiltonian.  Adding additional spin-spin interactions to the model  introduces dynamical bound pairs of $\mathbb{Z}_2$ fluxes in the ground state, with a characteristic binding length $l_F$, and characteristic time scale $\tau_F$.   Since the LL quantization essentially stems from the fact that our fermions perform cyclotron orbits whose size is set by $l_B$, with a characteristic time $\tau_B$, LL quantization is expected to persist as long as $l_F \gg l_B$ and $\tau_F \gg \tau_B$ such that the fermions can on average still perform cyclotron motion. 
Additionally, our finite-temperature data indicates that the LL peaks persist up to a small but finite {\it thermal} flux density of approximately $\rho_{\text{flux}} \approx 0.1 $.  
This suggest that the LL quantization may be more robust than this naive limit suggests: flux pairs will tend additionally to be bound to the sample's edges where the flux gap is reduced.  
  
Finally, we discuss the experimental feasibility of our proposal. 
Currently, different routes for the realization of $\alpha$-RuCl$_3$ thin films are being pursued, both via exfoliation of individual planes and by direct growth on substrates\,\cite{ziatdinov-17}. 
Strain could be generated either via spontaneously formed nano-bubbles, as for graphene~\cite{Levy10}, or by direct application of mechanical strain~\cite{Zhu15,guinea-10prb035408,Vozmediano10}. 
RuCl$_3$ flakes 
corresponding to our simulation would have diameters of $30$--$60$nm. For larger systems with $\beta=10$ the value of $s=0.04$ is expected to produce sharp LL peaks at a frequency that decays with the system size as in Fig. \ref{colorPlots}(a).
While standard Raman experiments average their $\mathbf{q}=0$ response over large areas of the sample, making observations challenging on small systems, it is possible to achieve a higher resolution by using a so called Raman microscope with spatial resolution sufficient for larger nano-bubbles.

Our conservative estimate of a cross-over temperature, together with an estimated Kitaev coupling of \mbox{$J\approx 100$K}~\cite{Sandilands2015,Banerjee2016}, yields detectable LL peaks (see Fig.\ref{colorPlots}(c)) at temperatures below roughly one Kelvin for the pure Kitaev model.  At these scales the separation of the LL peaks would be easily discernible with current energy resolution better than one meV in Raman experiments~\cite{Hackl2007}.  
However, if the KQSL phase can be achieved in these materials they will not be strictly at the Kitaev point and we expect the required energy and temperature scales to be further reduced.

{\paragraph{Summary.}
We have shown that Raman scattering can detect the quantized LLs that arise in strained honeycomb flakes of the Kitaev spin liquid, enabling a direct probe of the Dirac dispersion of the underlying Majorana fermions. Carrying out such an experiment on $\alpha$-RuCl$_3$ thin films is challenging --  but achievable -- with current technology.  Given that bulk $\alpha$-RuCl$_3$ is believed to be proximate to the KQSL phase~\cite{Banerjee2016}, it is not unreasonable to expect that thin films are free of the residual long-ranged magnetism because of their increased two-dimensionality, making the prospect of identifying a KQSL an exciting possibility in these systems.

\section*{Acknowledgements}
We acknowledge helpful discussions with Ken Burch, Alex Edelman and Arnab Banerjee. BP, FB and JK thank N.\ B.\ Perkins for collaboration on closely related work. SR thanks M.\ Vojta for discussions and collaborations on related projects.
BP was supported by the Torske Klubben Fellowhsip and the Doctoral Dissertation Fellowship. 
SR is supported by the German Science Foundation (DFG) through SFB 1143.
FJB is supported by NSF DMR-1352271 and Sloan FG-2015- 65927. 
The work of J.K. is supported by a Fellowship within the Postdoc-Program of the German Academic Exchange Service (DAAD). 


\bibliography{strain_refs}

\begin{thebibliography}{42}
\bibitem [{con()}]{conventions}%
  \BibitemOpen
  \href@noop {} {}\bibinfo {note} {Note that system size $r$ corresponds to
  $N-1$ in Ref.\,\cite{Rachel15}.}\BibitemShut {Stop}%
\end{thebibliography}

\newpage

\end{document}


\title{Supplemental Material for\\
	Majorana Landau Level Raman Spectroscopy}
		\author{Brent Perreault}
		\affiliation{\small School of Physics and Astronomy, University of Minnesota, Minneapolis, Minnesota 55455, USA}
		\author{Stephan Rachel}
		\affiliation{\small Institut f\"ur Theoretische Physik, Technische Universit\"at Dresden, 01062 Dresden, Germany}
		\affiliation{\small Department of Physics, Princeton University, Princeton, New Jersey 08544, USA}
		\author{F. J. Burnell}
		\affiliation{\small School of Physics and Astronomy, University of Minnesota, Minneapolis, Minnesota 55455, USA}
		\author{Johannes Knolle}
		\affiliation{\small Department of Physics, Cavendish Laboratory, JJ Thomson Avenue, Cambridge CB3 0HE, U.K.}

		
		\maketitle

\onecolumngrid
\appendix

\section{Finite Temperature Formalism} 
By the simple form of the canonical ensemble we can write 
\begin{align}\label{iflux}
I(\omega) \propto \sum_{\text{ flux patterns } M}  e^{-\beta E_0^M} I^M(\omega,\beta)
\end{align}
where $E_0^M$ is the energy of the lowest-energy state in the flux-configuration $M$, and $I^M(\omega,\beta)$ is the spectrum due to the band fermionic excitations in the flux background $M$, at temperature $\beta$. The proportionality constant is the partition function $Z = \sum_{\text{ flux patterns } M} e^{-\beta E_0^M} Z_M$ with $Z_M$ being the free fermion partition function in a given flux background. For a given flux-configuration $M$ the Hamiltonian can be written after Kitaev fermionization 
as
\begin{align}
\mathcal{H} =  \frac{1}{2}\sum_{\left<rr'\right>} J^{\alpha} u^\alpha_{\left<rr'\right>} i c_r c_{r'} \equiv  \frac{1}{2}\sum_{r,r'} H_{rr'} c_r c_{r'},
\end{align}
where $u^\alpha_{\left<rr'\right>} = i b^\alpha_r b^\alpha_{r'} = \pm 1$. 
In terms of these Majoranas $H$ has a chiral symmetry $S$ that flips sign of $c_r$ on one of two sublattices (so that $\{S,H\} = 0$). 
In the basis where $S$ is diagonal $H$ takes the block-off-diagonal form
\begin{align}
H = i\left(\begin{array}{cc}
0 & G \\
-G^\dag & 0
\end{array}\right).
\end{align}
In this case the  diagonalization of $H$ can be obtained from the singular value decomposition of $G$ \cite{Rachel15} (Supplemental Material). Given unitary $u$ and $v$ 
such that \mbox{$\ u^\dagger G v = \epsilon/4$}, then  
\begin{align}\label{U}
U = \frac{1}{\sqrt{2}}\left(\begin{array}{cc}
u & u \\
-iv & iv
\end{array}\right).
\end{align}
Now
\begin{align}
U^\dag H U = \Omega = \left[\begin{array}{cc}
\text{diag}(\vec{\epsilon}) & 0 \\
0 & -\text{diag}(\vec{\epsilon})
\end{array}\right],
\end{align}
with $\epsilon^\mu \geq 0$.
We can define operators $a_\lambda = U^\dag_{\lambda \lambda'} c_\lambda'$ for $\lambda = 1,...,n/2$ to get the set of $n/2$ fermionic quasiparticles, where $n$ is the number of unit cells,  \mbox{$\{a^\dagger_\lambda , a_{\lambda'} \}= \delta_{\lambda,\lambda'}$} so the Hamiltonian becomes
\begin{align}\label{Ha}
\mathcal{H} =  \frac{1}{2}\sum_{\lambda}\epsilon^\mu \left[2 a_\lambda^\dagger a_\lambda - 1 \right].
\end{align}
Therefore the excitation created by $a^\dagger_\mu$ has energy $\epsilon^\mu$.
These quasiparticles correspond to twice the positive-energy branch of the usual Dirac-fermions.

Following Ref. \cite{Nasu16} the Raman operator for the Kitaev model is given by 
\begin{align}
\mathcal{R} &= \sum_{\alpha = x,y,z} \sum_{\left<ij\right>_\alpha} \left({\boldsymbol \epsilon}_{\rm in} \cdot \mathbf{d}^\alpha \right) \left({\boldsymbol \epsilon}_{\rm out} \cdot \mathbf{d}^\alpha \right) J^\alpha S^\alpha S^\alpha  \\
& = \sum_{\left<rr'\right>} \left({\boldsymbol \epsilon}_{\rm in} \cdot \mathbf{d}^\alpha \right) \left({\boldsymbol \epsilon}_{\rm out} \cdot \mathbf{d}^\alpha \right) H_{rr'} c_r c_{r'} \\ 
&= \frac{1}{2}
\left(\begin{array}{c}
\mathbf{c}_A \\
\mathbf{c}_B 
\end{array}\right)^T
i\left(\begin{array}{cc}
A & B \\ -B^\dagger & A' 
\end{array}\right) \left(\begin{array}{c}
\mathbf{c}_A \\
\mathbf{c}_B 
\end{array}\right) \\
&= \frac{1}{2}
\left(\begin{array}{c}
\mathbf{a} \\
(\mathbf{a}^\dag)^T 
\end{array}\right)^\dagger
\left(\begin{array}{cc}
C & D \\ D^\dagger & -C 
\end{array}\right) \left(\begin{array}{c}
\mathbf{a}_{\lambda'} \\
(\mathbf{a}_{\lambda'}^\dag)^T 
\end{array}\right).
\end{align}
Here \mbox{$C = u^\dag B v + v^\dag B^\dag u$} and \mbox{$D = -u^\dag B v + v^\dag B^\dag u$} for a Raman operator that is symmetric w.r.t. swapping in and out polarizations and \mbox{$C = i\left( u^\dag A u + v^\dag A' v \right)$} 
and \mbox{$D = u^\dag A u - v^\dag A' v$} for an antisymmetric channel. $\mathbf{d}^\alpha$ is the bond vector for a bond of type $\alpha = x,y,z$, and ${\boldsymbol \epsilon}_{\rm in}$ and ${\boldsymbol \epsilon}_{\rm out}$ are the in and out light polarizations in the experimental setup. 
Then finally,
\begin{align}
I^M(\omega,\beta) \propto& \sum_{\lambda \lambda'} \left[ 2 |C_{\lambda \lambda'}|^2 f(\varepsilon_\lambda,\beta)[1-f(\varepsilon_{\lambda'},\beta)] \delta(\omega+\varepsilon_\lambda -\varepsilon_{\lambda'}) \right. \nonumber \\
&\left. +|D_{\lambda \lambda'}|^2 [1-f(\varepsilon_{\lambda},\beta)][1-f(\varepsilon_{\lambda'},\beta)] \delta(\omega-\varepsilon_\lambda -\varepsilon_{\lambda'})
\right] ,
\end{align}
where $B$ and therefore $C$ and $D$ depend on the gauge chosen for each flux configuration $M$, and $f(\varepsilon,\beta)$ is the familiar Fermi distribution. The gauge chosen in each case is obtained by creating the shortest-distance gauge string from the nearest edge plaquette for every flux in the desired configuration. Of course, all measurable quantities discussed here are independent of the gauge choice \cite{Perreault-thesis}. 

\begin{figure}[b!]
	\centering
	\includegraphics[width=0.43\linewidth]{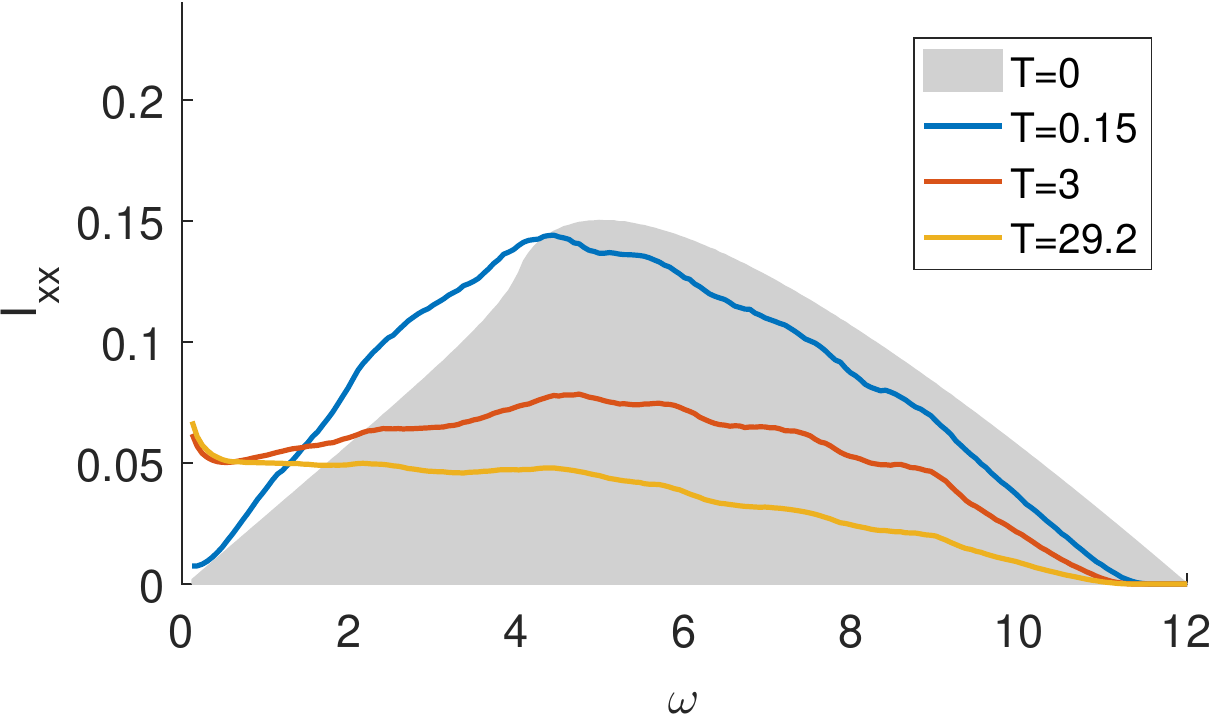} 
	\includegraphics[width=0.43\linewidth]{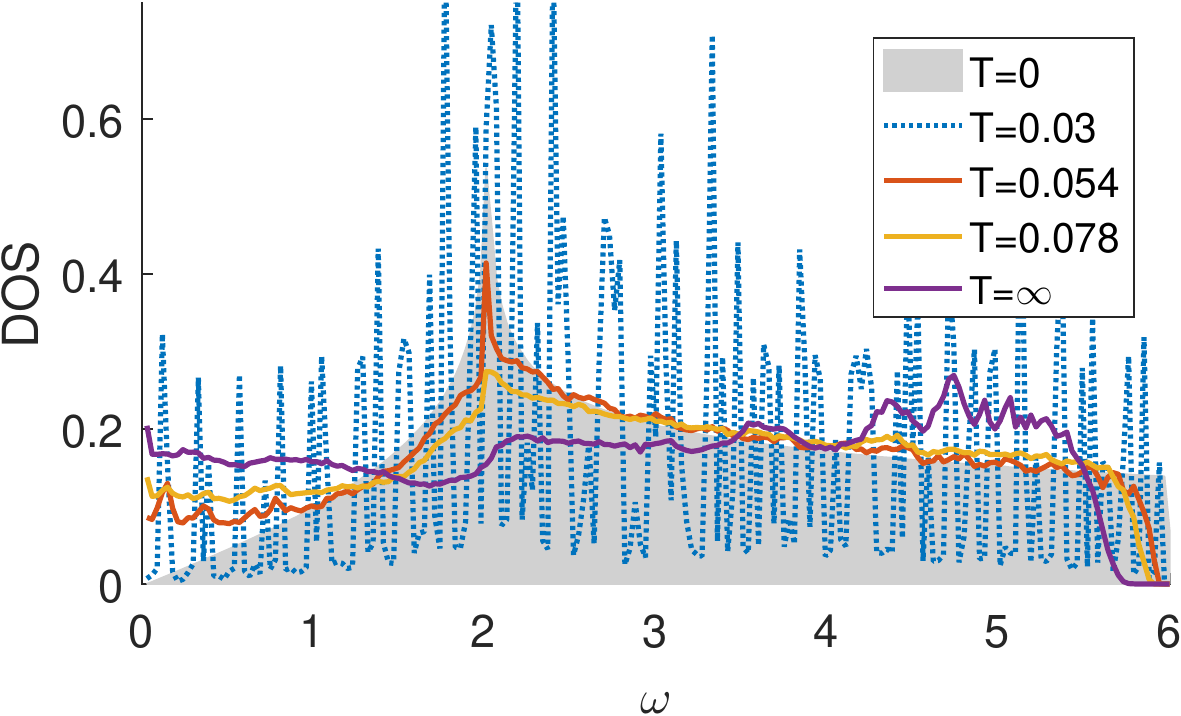} 
	\caption{Plots of unstrained Raman and DOS to compare with Ref. \cite{Nasu16} Fig. 3(a) and Ref. \cite{Nasu15} Fig. 3(a). resp. The comparisons appear quite favorable, although the present study did not involve the superlattice treatment that allowed Ref. \cite{Nasu15} to suppress the finite-size effects that are large in our plot of the DOS. The system sizes used were $ r = 11$ and $r=7$ chosen to most closely mimic the ones used in those references. }
	\label{comparePlots}
\end{figure}

We evaluate the sum (\ref{iflux}) using Markov Chain Monte Carlo (MCMC) of the usual kind, which we review now. The Markov chain is a series of flux configurations that represent ones taken from the Boltzmann distribution for a fixed temperature. The chain is built by proposing steps that either (1) flip a random plaquette, (2) move a single plaquette to another empty space, (3) flip a plaquette that is a neighbor or next-nearest-neighbor (NNN) of one of the current fluxes (including the fluxes themselves) and (4) shuffle the fluxes to a random pattern with the same number. If a proposed flux configuration is accepted, it is appended to the chain, otherwise it is denied, and the current state is appended to the chain again. To recreate the correct distribution, the probability of acceptance is set to be
\begin{align}
\mathcal{P}(p'|p) = \min \left( e^{-(E_{p'}^0 - E_p^0)} \frac{\mathcal{Q}(p|p')}{\mathcal{Q}(p'|p)} , 1\right),
\end{align}
where $\mathcal{Q}(p'|p)$ is the probability of proposing configuration $p'$ given that you are in configuration $p$. For proposals (1), (2) and (4) $\mathcal{Q}(p'|p) = \mathcal{Q}(p|p')$, but for the case of NNNs (3) the new configuration $p'$ can have a different number of plaquettes that are NNNs of fluxes, making the probability of flipping the same plaquette slightly different. In this case the term can be interpreted as a relative entropy of possible proposals before and after the given proposal.

The error in the result is analyzed using Geyer's initial convex sequence estimator (ICSE) $g$ for the variance along a Markov chain.\cite{Geyer92} This measure takes into account the local correlation between events, thereby producing a reliable, tight upper-bound estimate for the sample variance given the data. Then the error in the mean for a given observable can be computed as $\sqrt{(g(X)/N)}$, where $X$ is the Markov chain of that observable, N is the length of the chain. The ICSE was computed using the initseq function implemented in R by Geyer in the mcmc contributed package on CRAN, translated to Matlab for this project. In addition, the autocorrelation time, or the number of steps/proposals required to achieve an effective independent sample can be estimated by $\tau(X) = g(X)/\text{Var}(X)$ where $\text{Var}$ computes the usual variance of the values in the chain $X$.

The algorithm was run until either the mean error relative to the mean value of each observable was below $1/65$, or 50,000 proposals steps were made. The mean expected error relative to the mean value for a given temperature was below 2\% in all cases. 
Comparison with previous work at zero strain is made in Fig. \ref{comparePlots}. 

\begin{figure}[t!]
	\centering
	\includegraphics[width=\linewidth]{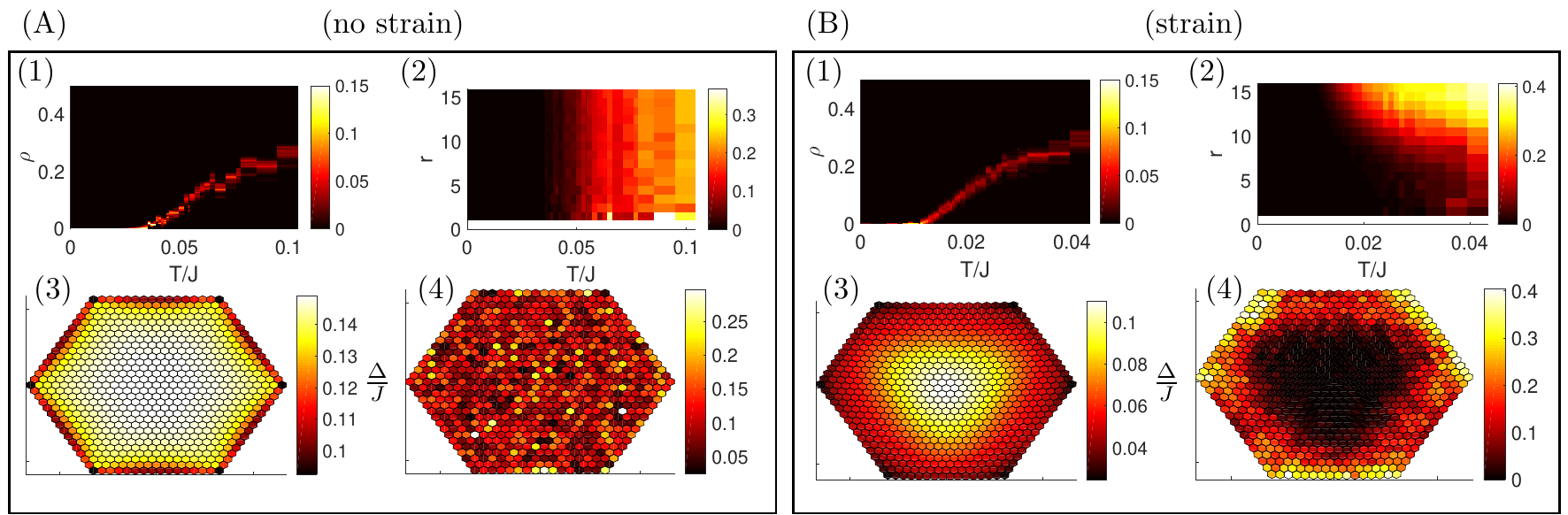}
	\caption{The nature of the crossover out of the KQSL phase is demonstrated for (A) an unstrained flake and (B) a strained flake with $s=0.04$. For uniform non-interacting fluxes the flux density in (1) is expected to follow a Fermi-Dirac distribution at energy $(\ln 2)\Delta$ where $\Delta$ is the flux gap \cite{Perreault-thesis}. (2) shows the flux density resolved over the radial coordinate as a function of temperature. (3) plots the single-flux gap as a function of position on the honeycomb flake \cite{Rachel15}. (4) give the expectation value of the flux operator as a function of position for one particular run at temperatures $T = 0.057 J$ in (A,4) and $T = 0.020J$ in (B,4), both corresponding to the critical flux density of the strained flake $\rho_c \approx 0.12$. Although there is significant noise in (4) we see that the flux distribution is largely random without strain (A,4) and it becomes focused away from the center under stain. This effect is responsible for the appreciable critical flux density $\rho_c \approx 0.12$ for Landau levels in the strained flake, for which Landau orbits can form within the `clean' center region.}
	\label{thermodynamics}	
\end{figure}

Finally, in Fig. \ref{thermodynamics} we compare the character of the finite temperature crossover at zero strain to that of the strained system.  The role of the spatial flux-gap anisotropy (shown in panel (3) of each subfigure) is apparent: at zero strain where the flux gap is essentially uniform away from the sample boundaries, the crossover entails a proliferation of fluxes with a uniform spatial probability distribution; the temperature scale for this cross-over is set by the bulk flux gap.  In the strained system, the flux gap depends strongly on position within the honeycomb flake, and fluxes first proliferate predominantly near the sample boundaries, only reaching a uniform distribution at significantly higher temperature.  The temperature scale fo this initial increase in flux density is set by the flux gap near the flake's boundary, which is considerably lower than that of the un-strained system.

\bibliography{strain_refs}